%% file: main.tex
\documentclass[11pt]{article}
\pdfoutput=1

\usepackage[T1]{fontenc}              
\usepackage[latin9]{inputenc}         

\usepackage[a4paper]{geometry}        

\usepackage{url}

\usepackage{booktabs}

\usepackage{graphicx} 
\usepackage{amssymb,amsfonts,amsmath}

\usepackage{natbib}
\usepackage{mathpazo}
\usepackage{microtype}

\usepackage{color}

\input{abbreviations}

\begin{document}

\date{1 June 2012; final revision 25 October 2012}

\title{
A novel algorithm for simultaneous SNP selection in high-dimensional genome-wide association studies
}

\author{Verena Zuber
      \thanks{Institute for Medical Informatics,
      Statistics and Epidemiology,
      University of Leipzig,
      H\"artelstr. 16--18,
      D-04107 Leipzig, Germany} 
{}, 
A. Pedro Duarte Silva
\thanks{Faculdade de Economia e Gest\~{a}o \& CEGE, 
        Catholic University of Portugal,
       Rua Diogo Botelho 1327, 
      4169-005 Porto, Portugal} 
{} and Korbinian Strimmer \footnotemark[1]
}

\maketitle
\begin{abstract}
\input{0-abstract}

\end{abstract}

\newpage

\input{1-introduction}

\input{2-methods}

\input{3-results}

\input{4-discussion}

\input{5-acknowledgments}


\bibliographystyle{apalike}
\bibliography{preamble,econ,genome,stats,array,sysbio,misc,molevol,med,entropy}

\end{document}

%% file: abbreviations.tex
\urldef\cranurl\path{http://cran.r-project.org/}

\newcommand\bI{\boldsymbol I}
\newcommand\bM{\boldsymbol M}

\newcommand\bP{\boldsymbol P}
\newcommand\bR{\boldsymbol R}

\newcommand\btau{\boldsymbol \tau}

\newcommand\bU{\boldsymbol U}

\newcommand\bX{\boldsymbol X}

\newcommand{\figcite}[1]{Fig.~\textbf{\ref{#1}}}
\newcommand{\eqcite}[1]{Eq.~\textbf{\ref{#1}}}
\newcommand{\tabcite}[1]{Tab.~\textbf{\ref{#1}}}

%% file: 0-abstract.tex
\noindent\textbf{Background:}
Identification of causal SNPs in most genome wide
association studies  relies on approaches 
that consider each SNP individually.
However, there is a strong correlation structure among SNPs
that need to be taken into account. Hence, increasingly modern 
computationally expensive regression methods are employed
for SNP selection that consider all markers simultaneously
and thus incorporate dependencies among SNPs.

\noindent\textbf{Results:}
We develop a novel multivariate algorithm for large scale
SNP selection using CAR score regression, a promising new approach
for prioritizing biomarkers.  
Specifically, we propose a computationally efficient procedure for shrinkage
estimation of CAR scores from  high-dimensional data.
Subsequently, we conduct a comprehensive comparison study including five advanced 
regression approaches (boosting, lasso, NEG, MCP, and CAR score) 
and a univariate approach 
(marginal correlation) to determine the effectiveness in finding  true causal SNPs.

\noindent\textbf{Conclusions:}
Simultaneous SNP selection is a challenging task. We demonstrate that our
CAR score-based algorithm
consistently outperforms all competing approaches, both uni- and multivariate,
in terms of correctly recovered causal SNPs and SNP ranking.
An R package implementing the approach as well as R code to reproduce the complete
study presented here is available from \url{http://strimmerlab.org/software/care/}.

%% file: 1-introduction.tex
\section*{Background}

Genome-wide associations studies (GWAS) are now routinely conducted to 
search for genetic factors indicative of or even causally linked to disease.
Typically, the aim of such a study is to identify a small subset of 
single nucleotide polymorphisms (SNPs) associated with a phenotype of interest.
From an analysis point of view the screening for relevant biomarkers is best
cast as a problem of statistical variable selection.  In GWAS variable selection
is very challenging as the full set of SNPs is often very large while both the
effect of each potentially causal SNP as well as their number is very 
small \citep[e.g.][]{HB2008,AC2010,GS2011}.

To date, most GWAS are based on single-SNP  analyzes
where each SNP is considered independently of all others
and  association with the phenotype is computed using a univariate
test statistic such as variants of the $t$-score, the ATT
statistic \citep{Arm1955} or marginal correlation \citep{Fou2009}.
The advantage of this approach is that it is computationally 
inexpensive. However, it implicitly assumes complete independence of markers
and thus ignores the correlation structure among SNPs, e.g., due to linkage or 
interaction among SNPs.

In order to increase statistical efficiency and to exploit the
correlation among predictive SNPs several authors  have recently 
started to investigate
simultaneous SNP selection using fully multivariate approaches.
This was pioneered for GWAS in the seminal paper of \citet{HB2008}
that introduced the NEG regression model, a shrinkage-based approach
to select relevant SNPs.  A related approach is LASSO regression
that was employed to GWAS by
\citet{WC+2009}, MCP regression \citep{AC2010}, and
Bayesian variable selection regression 
\citep{GS2011}.  Another promising multivariate approach advocated for
 high-dimensional variable selection is boosting \citep{HB2006}
but this has not yet been investigated for GWAS.

Recently, to address the problem of variable importance and
selection under correlation in genomics, we have introduced two
novel statistics, the correlation-adjusted $t$-score (CAT score)
and the correlation-adjusted
marginal correlation (CAR score), see \citet{ZS09,ZS2011}.
These two measures are multivariate generalizations
of the standard  univariate test statistics that take the
correlation among variables explicitly into account and lead
to improved rankings of markers as has been shown for data from 
transcriptomics and metabolomics.
However, application of CAT and CAR scores has
so far been restricted to medium to large dimensional settings only
as  computing these scores involves the calculation of the
inverse matrix square root of the correlation matrix, which is prohibitively
expensive in high dimensions.  Thus, for SNP analyzes
further computational economies are needed.

Here, we develop a novel multivariate algorithm for large scale
SNP selection using CAR score regression.
Specifically, we propose a computationally efficient procedure that allows for shrinkage
estimation of CAR scores even for very high-dimensional data sets.
Subsequently, we conduct a systematic comparison of state-of-the-art simultaneous
SNP selection procedures  using data from the GAW17 consortium \citep{AD+2011}.
These data are particularly suited for investigating relative performance 
 as  the true causal SNPs are known.
Finally, we demonstrate that SNP rankings based on correlation-adjusted statistics 
consistently outperform all investigated competing approaches, both uni- and multivariate.

%% file: 2-methods.tex

\section*{Methods}

\subsection*{Univariate ranking of SNPs}

The basic setup we consider here is a linear regression model for
a set of $d$ predictors  $\bX = \{X_1, \ldots, X_d\}$ and
a metric or binary response variable $Y$.
In GWAS the covariates $\bX$ are given by the genotype 
and the response $Y$ is the phenotype or trait  of interest. 
The correlation matrix among the predicting variables has size
 $d \times d$ and is denoted by 
$\bP$ (capital ``rho''). 
 The vector of marginal correlations
$\bP_{\bX Y} = (\rho_{X_1 Y}, \ldots, \rho_{X_d Y})^T$ contains
the correlations between a metric response and each individual SNP.
Similarly, for binary response the $t$-score vector
$\btau = (\tau_1, \ldots, \tau_d)^T$ contains the $t$-scores computed
for each variable.

If there is no correlation among SNPs (i.e. 
$\bP = \bI_d$) the $t$-scores $\btau$ provide an optimal ranking of SNPs
in terms of predicting a binary $Y$ \citep{Efr09}. Likewise, for 
metric response the marginal correlations 
lead to an optimal ordering  \citep{FL2008}.
Moreover, in the
absence of SNP-SNP correlation the squared values of
the ranking statistics (squared $t$-score, squared marginal
correlation) are useful measures of variable importance,
adding up to Hotelling's $T^2$ and the squared multiple correlation coefficient
$R^2$, respectively.

\subsection*{CAT and CAR score}

In many important settings the correlations $\bP$ do not vanish but
rather represent additional structure relating the predictors.
In the case of SNPs the correlation may be rather large, 
e.g. due to linkage effects \citep{AKS2002}.
Thus, both for variable ranking and for assigning variable importance 
it can be essential to take the correlation between covariates into account.

To this end we have proposed a simple modification of the $t$-statistic
and marginal correlations.  In \citet{ZS09} we have introduced the CAT
score (correlation-adjusted $t$-score) that is defined as
\begin{equation}
\btau^{\text{adj}}=\bP^{-1/2} \btau 
\label{eq:catscore}
\end{equation}
where $\bP^{-1/2}$ is the inverse of the matrix square-root of $\bP$.
The vector $\btau^{\text{adj}}$ contains the adjusted $t$-scores  which measure 
the influence of each predictor on $Y$ after simultaneously removing the effect
of all other variables.   The squared CAT score may thus be used as measure
of variable importance. Unlike squared $t$-scores they sum up to Hotelling's $T^2$ even in 
the presence of correlation,
$$
(\btau^{\text{adj}})^T \btau^{\text{adj}} = \btau^T \bP^{-1} \btau = T^2 .
$$
Correspondingly, in \citet{ZS2011} we investigated a correlation-adjusted
marginal correlations (CAR scores)
\begin{equation}
\bP_{\bX Y}^{\text{adj}}=\bP^{-1/2} \bP_{\bX Y} \,.
\label{eq:carscore}
\end{equation}
 The squared CAR scores
sum up to the squared multiple correlation coefficient
$$
(\bP_{\bX Y}^{\text{adj}})^T \bP_{\bX Y}^{\text{adj}} = \bP_{Y \bX} \bP^{-1} \bP_{\bX Y} = R^2 \, ,
$$
also known as coefficient of determination or proportion of variance explained.
Because of this decomposition property CAT and CAR scores allow to assign importance
not only to individual SNPs but also to groups of SNPs.  
Moreover, both CAT and CAR score share a grouping property that leads
to similar scores for highly correlated SNPs. In addition they
protect against antagonistic SNPs, i.e. if two SNPs are highly correlated
and one has a protective and the other a risk effect, then both SNPs 
are assigned low scores.

For model selection using CAT and CAR scores, i.e. for identification of those SNPs that do not contribute to predict
the response $Y$, we use a simple thresholding procedure with the critical
 threshold obtained by controlling local false discovery rates \citep{KS2012}.

In previous work we have shown for synthetic data as well as for data from
metabolomic and gene expression experiments that CAT and CAR scores
are effective multivariate criteria for obtaining compact yet highly predictive feature sets.
Independently, in the study of \citet{AT2012} it was also found that CAT scores result in favorable orderings of variables.

However,  with increasing dimension $d$ the correlation matrix $\bP$ becomes prohibitively
large both to compute and to handle effectively. As a result, in high dimensions
direct calculation of CAT and CAR scores using
\eqcite{eq:catscore} and \eqcite{eq:carscore} is not possible.
Thus, for application in high-dimensional data such as from GWAS an alternative means of computation must be developed.

\subsection*{Computationally efficient calculation of shrinkage estimators
of CAT and CAR scores}

If the number of observations $n$ is smaller than the number of variables
$d$ we need to employ a regularized estimate for the correlation matrix $\bP$.
A simple  shrinkage estimator $\bR$ for $\bP$
is given by  
$$
\bR =  \lambda \bI_d + (1-\lambda) \bR_{\text{empirical}} 
$$
where $\bR_{\text{empirical}}$ is the empirical non-regularized correlation matrix and
$\lambda$ is a shrinkage intensity \citep[e.g.][]{SS05c}.
Using  computational economies akin to those discussed in \citet{HT04}
we now show that  computation of  $\bR^{-1/2}$ and subsequent calculation
of estimates of  CAT and CAR scores 
can be done in a computationally highly effective way, even when
direct computation of CAT and CAR scores via \eqcite{eq:catscore} and \eqcite{eq:carscore}
is infeasible.

Using singular value decomposition the empirical correlation matrix
can be written 
$\bR_{\text{empirical}} = \lambda/(1-\lambda) \bU \bM \bU^T$
where $\bM$ is  positive definite matrix of 
size $m \times m$, $\bU$ an orthonormal matrix of size $d \times m$, 
and  $m = \text{rank}(\bR_{\text{empirical}}) << d$.
This simplifies the shrinkage estimator to
$$
\bR =  \lambda (\bI_d + \bU \bM \bU^T ) \,.
$$
Following \citet{ZS09} we then compute the $\alpha$-th matrix power of $\bR$
using
$$
 \bR^\alpha =   \lambda^{\alpha} (\bI_d - \underbrace{\bU}_{d \times m}  (\bI_m-(\underbrace{\bI_m + \bM}_{m \times m})^\alpha) \underbrace{\bU^T}_{m \times d} ) \,.
$$
This implies we only have to compute the matrix power of the 
$m \times m$ matrix $\bI_m + \bM$ 
to obtain $\bR^\alpha$.  Moreover, for efficiently calculating  CAT and CAR scores it is crucial to note that
it is not at all necessary neither to store or to compute the full $d \times d$ sized matrix $\bR ^{-1/2}$
 as  
\begin{equation}
\begin{split}
\bR_{\bX Y}^{\text{adj}} & = \bR^{-1/2} \bR_{\bX Y}\\ 
& =  \lambda^{-1/2} (\bI_d -\bU (\bI_m-(\bI_m + \bM)^{-1/2}) \bU^T )   \bR_{\bX Y} \\
& = \lambda^{-1/2} ( \underbrace{\bR_{\bX Y}}_{d \times 1} - \underbrace{\bU (\bI_m-(\bI_m + \bM)^{-1/2}}_{d \times m} ) (\underbrace{\bU^T  \bR_{\bX Y}}_{m \times 1} )) \,.
\end{split}
\label{eq:fastcarscore}
\end{equation}
Consequently, \eqcite{eq:fastcarscore} allows to obtain shrinkage estimates of CAT and CAR scores effectively even in high dimensions
as none of the matrices employed in \eqcite{eq:fastcarscore} is larger than $d \times m$, and most are even smaller
($d \times 1$ or $m \times 1$), all without actually computing the shrinkage correlation matrix $\bR$.

%% file: 3-results.tex

\section*{Results and Discussion}

We now compare the proposed CAR score approach to simultaneous SNP selection 
with competing methods and determine its  effectiveness in finding true
causal SNPs.

For this purpose we use the mini-exome data set compiled for
the GAW17 workshop held 13-16 October 2010 in Boston 
(\url{http://www.gaworkshop.org/gaw17/}).  This data set 
is a combination of real sequence data  and simulated synthetic phenotypes,
where the true causal SNPs are known.  In our study we investigate univariate
ranking by marginal correlation and five multivariate approaches.

In order to facilitate replication of our results we provide complete R code \citep{RPROJECT}.
Our R package ``care'' implements the developed algorithm.  Moreover, we offer R scripts 
covering all analysis steps from preprocessing the raw data to plotting of figures
at \url{http://strimmerlab.org/software/care/}.  The data are publicly
available from the GAW consortium, see
\url{http://www.gaworkshop.org/gaw17/data.html} for details.

\subsection*{GAW 17 unrelated data}

The compilation and simulation of phenotypes for the GAW17 mini-exome
data set is described in detail in \citet{AD+2011}.
We focus here on the GAW 17 unrelated data with metric 
phenotypes Q1, Q2, and Q4.
The corresponding sequence data matrix contains information on 24,487 SNPs
for  $n=697$ individuals. For each phenotype there are $B=200$ simulations. 
By construction, phenotype
Q1 has a  residual heritability of 0.44
and is influenced by 39 SNPs in 9 genes,
whereas  Q2 has a lower residual heritability of 0.29
and is influenced by 72 SNPs in 13 genes.
This suggests that discovery of 
true causal SNPs should be less challenging for Q1 than for Q2.
Phenotype Q4 has a heritability of 0.70 but none of it is due to SNPs
contained in the present data set.

\subsection*{Preprocessing}

In the preprocessing of the sequences
we first recoded the alleles in the raw data  into 0, 1, 2 
assuming an additive effects model.  Second, we standardized the data matrix
 to column mean zero and column variance 1.   
Subsequently, we removed duplicate 
predictors so that 15,076 unique SNPs remained.
The set of true causal SNPs for both Q1 and Q2 also contains
each a duplicate, reducing the number of true unique SNPs to
38 and 71.
Finally, we further filtered out synonymous SNPs,
as we are interested only in non-synonymous mutations.
The resulting predictor matrix $\bX$ is of size $697 \times 8,020$,
i.e.\  $d=8,020$ unique non-synonymous SNPs are simultaneously considered for
selection.

For preprocessing the response variables Q1, Q2, and Q4 we removed 
the influence of the three non-genetic covariates sex, age, and smoking
by linear regression.  The resulting residuals were standardized
to mean zero and variance 1 which yielded $B=200$ response vectors
$y_1^{(b)}$, $y_2^{(b)}$, and $y_4^{(b)}$, where $b \in 1,...,B$, each of size $697 \times 1$.

\subsection*{SNP selection methods included in the comparison study}

\nocite{BH2011}
\nocite{FHT2010}

\begin{table}[h]
\caption{Software used in the comparison study. The R packages are available
from the R software archive CRAN at \cranurl.}
\centering
\begin{tabular}{lll}
\toprule
Method         & Software & Reference   \\
\midrule
CAR   &  R package {\tt care} & \citet{ZS2011} \\
COR   &  R package {\tt care} & \citet{ZS2011} \\
NEG   & {\tt HLasso} program  & \citet{HB2008} \\
MCP   & R package {\tt ncvreg} & \citet{BH2011} \\
BOOST & R package {\tt mboost} & \citet{HB2006} \\
LASSO & R package {\tt glmnet} & \citet{FHT2010} \\
\bottomrule
\end{tabular}
\label{tab:software}\\

\end{table}

For each of the $B=200$ response vectors for Q1, Q2, and Q4 we computed a regression
model including all $d=8,020$ SNPs as potential predictors.
Following \citet{AC2010} we focused on regularized regression approaches.
Specifically, we used the following five methods, all of which
have been shown to be powerful tools for variable selection in 
large-scale regression settings:
\begin{itemize}
\item CAR: variable ranking by shrinkage CAR scores \citep{ZS2011},
\item NEG: regression with normal exponential gamma (NEG) prior  \citep{HB2008},
\item MCP: regression with MCP penalty \citep{Zha2010},
\item BOOST: boosting \citep{Sha1990}, and
\item LASSO: lasso regression \citep{Tib96}.
\end{itemize}
The corresponding software implementations are listed in \tabcite{tab:software}.
As a reference for comparison  we additionally included two baseline 
methods:
\begin{itemize}
\item COR: univariate SNP ranking by marginal correlation, and
\item RND: random ordering of all SNPs.
\end{itemize}
All methods except CAR and COR combine regularization with variable selection.
Thus, for determining model sizes for CAR scores and COR we 
adaptively estimated a threshold from the data
using a local FDR cutoff of 0.5 as recommended in \citet{KS2012}.
In settings with rare
and weak features this particular choice coincides with the so-called
``higher criticism'' threshold that has shown to be powerful
for signal identification in classification \citep[e.g.,][]{DJ08,DJ2009,Dua2011}.
For computing the FDR values we employed the R package {\tt fdrtool} \citep{Str08b,Str08c}.

Generally, all software were 
run with default settings.
The regularization parameters required by the NEG, MCP, BOOST and CAR approaches were set to fixed values 
optimizing the overall performance of each method. Specifically, for CAR and MCP we employed
$\lambda=0.1$, for BOOST $\nu=0.1$, and for NEG $\lambda=85$. 
For LASSO we used the built-in cross-validation routines.

\subsection*{Relative performance of investigated methods}

\begin{figure*}[hp]
\begin{center}
\includegraphics[width=15cm]{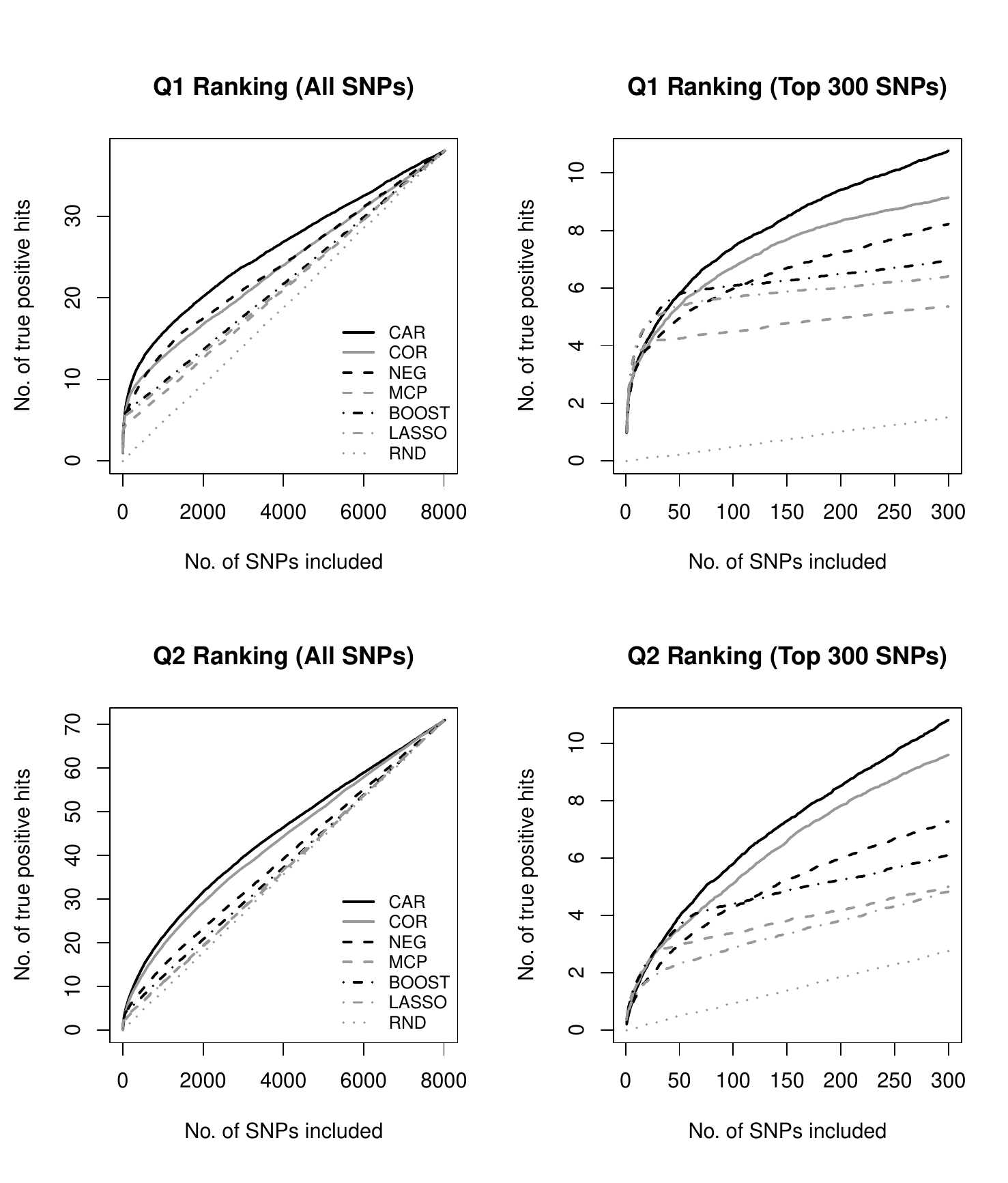}
\caption{Average true positives resulting from SNP rankings of the investigated
approaches for phenotype Q1 (top row) and Q2 (bottom row). For Q1 there
are 38 true SNPs and for Q2 71 true SNPs.\label{fig:rankingQ1Q2}}
\end{center}
\end{figure*}

\begin{table}[h]
\caption{Median model sizes and the corresponding
interquartile ranges (IQR) as well as the average true positives for 
phenotypes Q1 and Q2 for all investigated methods
summarized across the 200 repetitions (first three columns).
For comparison, the last three columns show the average true positives
at the specified model size for CAR, COR and RND.
The best performing method is
shown in bold, the second best in italic.}
\centering
\begin{tabular}{lrrr r rrr}
\toprule
&\multicolumn{3}{c}{Results} & & \multicolumn{3}{c}{Comparisons} \\
\midrule
&Method         & Model Size     & TP & & TP & TP &TP   \\
&               & Median (IQR)   & Method & & CAR & COR & RND \\
\midrule
Q1 \\
&CAR       &   51 (53)  & \bf  5.85 & & \bf 5.85  & \it 5.42 & 0.23 \\
&COR       &  176 (108) & \it  8.06 & & \bf 8.99  & \it 8.06 & 0.88 \\
&NEG       & 1390 (118) & \it 15.31 & & \bf 17.57  &    14.38 & 6.60 \\
&MCP       &   20 (5)   & \it  4.11 & & \bf  4.19  &     3.95 & 0.12 \\
&BOOST     &   53 (5)   & \it  5.84 & & \bf  5.91  &     5.50 & 0.25 \\
&LASSO     &   37 (31)  & \it  5.19 & & \bf  5.21  &     4.89 & 0.18 \\
\midrule
Q2 \\
&CAR     &   31 (38)  & \bf 2.93 & &  \bf 2.93 &  \it 2.85 &  0.29 \\
&COR     &    1 (7)   & \bf 0.38 & &  \it 0.21 &  \bf 0.38 &  0.00 \\
&NEG     & 1632 (755) &    20.21 & & \bf 28.08 & \it 25.90 & 14.50 \\
&MCP     &   29 (5)   &     2.75 & &  \bf 2.82 &  \it 2.76 &  0.28 \\
&BOOST   &   59 (6)   & \it 3.92 & &  \bf 4.34 &      3.82 &  0.59 \\
&LASSO   &   15 (36)  &     1.50 & &  \it 1.88 &  \bf 1.97 &  0.14 \\
\bottomrule
\end{tabular}
\label{tab:sizetpQ1Q2}\\

\end{table}

\begin{table}[h]
\caption{Median model sizes and the corresponding
interquartile ranges (IQR) for phenotype Q4.}
\centering
\begin{tabular}{l rrrr rrr}
\toprule
Q4 \\
&Model Size  & CAR &  COR  &   NEG  &  MCP &   BOOST &  LASSO \\
&    Median  &  34 &    0  &  1900  &   27 &      59 &      1 \\
&    IQR     &  40 &    1  &  2713  &    4 &       6 &      6 \\
\bottomrule
\end{tabular}
\label{tab:sizeQ4}\\

\end{table}

The aim of this study is to compare simultaneous SNP selection methods
with regard to their ability to discover the true known SNPs.  
For this purpose we investigated the respective SNP rankings and the corresponding
true positives, the size of the selected models, and 
the variability across
the 200 repetitions.

In \figcite{fig:rankingQ1Q2} and the associated \tabcite{tab:sizetpQ1Q2} 
we compare the effectiveness of SNP rankings for phenotypes Q1 and Q2.
For Q1 all methods uniformly outperform marginal correlation, i.e. at the model size
determined by each procedure the number of true positives is larger than
that for  marginal correlations at the same cutoff.  Thus, for Q1 all multivariate
SNP selection approaches improve over univariate selection. Moreover, as can be seen from
 \figcite{fig:rankingQ1Q2} (top row) and \tabcite{tab:sizetpQ1Q2} 
for small numbers of included SNPs all methods perform similarly but 
starting from model size of 50 SNPs  CAR scores lead to a  better ranking
 in terms of true positives  than all other competing
approaches.   For the more challenging phenotype Q2 the situation is similar.  CAR scores
almost always provide the most effective ranking (see lower part of \tabcite{tab:sizetpQ1Q2})
but intriguingly for this phenotype it is also the only multivariate method that improves
over marginal correlation.   

In \tabcite{tab:sizetpQ1Q2} we also list the median model
sizes for each regression approach.  LASSO and MCP generally lead to small
numbers of selected SNPs (less than 40), BOOSTING, CAR and COR variable sets are medium sized and
NEG chooses a very large number of SNPs. Note the variability in the estimated
model sizes as quantified by the corresponding interquartile ranges (IQR)
is largest in the methods that estimate the threshold adaptively from the data (CAR, COR, LASSO) whereas it is smallest for those methods where we
used a fixed regularization parameter (NEG, MCP, BOOST).
Finally, in \tabcite{tab:sizeQ4} the model sizes and IQR for phenotype
Q4 is shown for the investigated methods.  Here, COR and LASSO lead to the smallest model sizes and thus the smallest number of false positives,
with the MCP and CAR methods being the runners-up.

\begin{figure*}[p]
\begin{center}
\includegraphics[width=10cm]{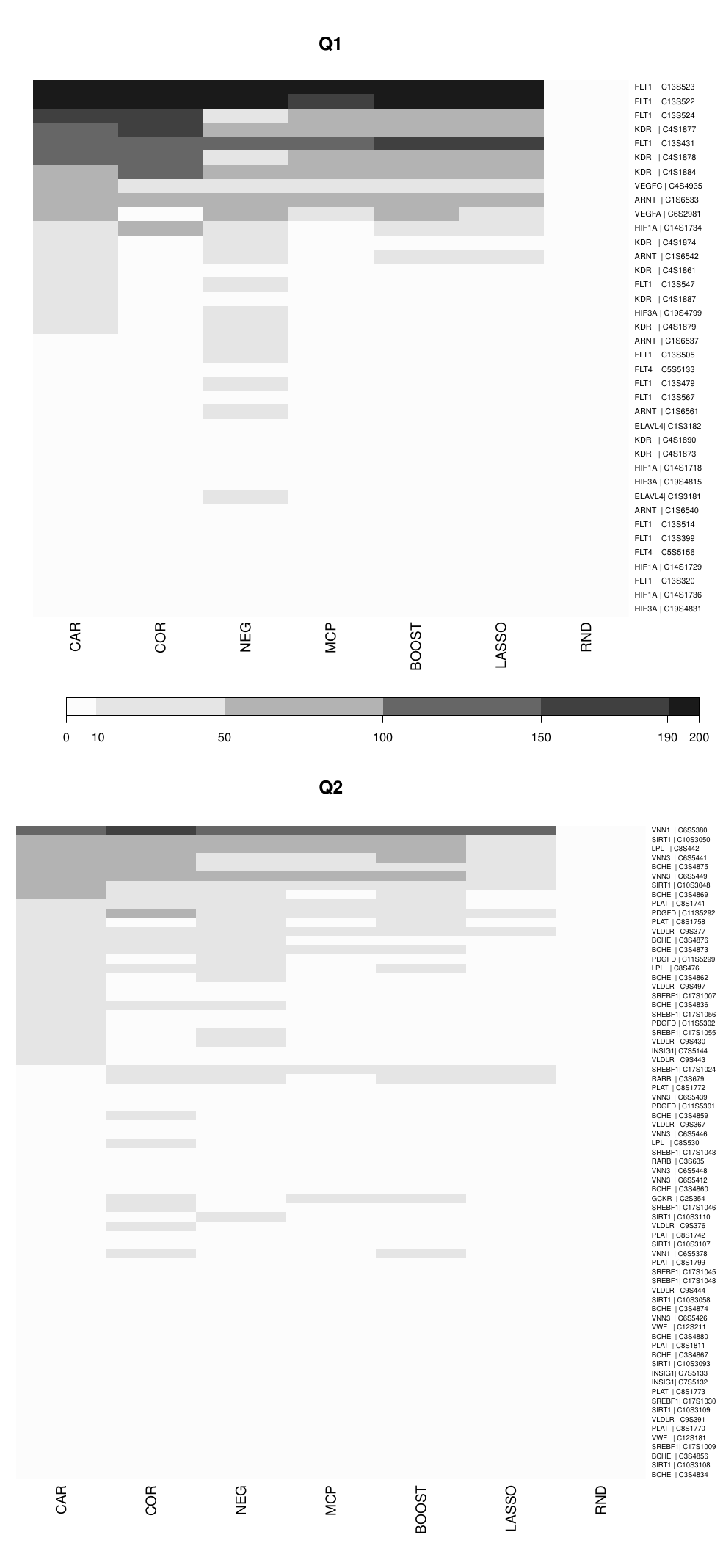}
\caption{Frequency of occurrence of each true SNP among the top 100 SNPs selected by each approach
for phenotype Q1 (top row) and for Q2 (lower row) for the 200 repetitions. Note
that the SNPs are ordered according to the first column.\label{fig:heatmapQ1Q2}}
\end{center}
\end{figure*}

In further investigation of these results we identified the actual true SNPs recovered by
each SNP selection approach.   Specifically, we counted which of the 38 respectively 71 
true causal SNPs for Q1 and Q2 were found among the first 100 top ranking SNPs
using the 200 repetitions available for each phenotype.
The result is shown as a heatmap in \figcite{fig:heatmapQ1Q2}
and visualizes the relative difficulty of recovering the individual causal SNPs.
In Q1, there are two SNPs on top of the heatmap that are consistently 
detected by all methods.  Then, there is a large block primarily
recovered by CAR score and correlation, but not by the other approaches.
Finally, there are some moderate detections only in CAR scores and NEG regression.
Half of the true positives are hardly discovered by any method. 
The comparison with randomly ordered SNPs (column RND) shows that those SNPs
only appear by chance. 
For Q2, there is only a single SNP that is consistently  included in all models.
As in Q1, it is followed by a small group of detections most prominent in CAR 
score and correlation.  Finally, there are some moderate findings for 
both, the CAR score and NEG, and some only for correlation.
In addition, hierarchical clustering of the columns (methods) in this heatmap (tree not shown in figure)
 reveals a basic similarity pattern among the methods: 
CAR and COR cluster together, NEG and MCP regression form another cluster, and LASSO and BOOST
are grouped together.

\begin{table}[h]
\caption{True SNPs found among the top 100 SNPs identified by CAR scores in at least 50 of the 200 repetitions for Q1 and Q2.
The last column shows the average absolute correlation among all SNPs for Q1 and Q2 as well as the average
absolute correlation for the SNPs belonging to one gene.}
\centering
\begin{tabular}{llrrrr}
\toprule
&SNP         & Frequency & MAF & BETA & Correlation \\
\midrule
Q1 &&&&& 0.014 \\
& ARNT  $|$ C1S6533   &  88 & 0.011478 & 0.56190 & \\
& FLT1  $|$ C13S431   & 110 & 0.017217 & 0.74136 & 0.147 \\
& FLT1  $|$ C13S522   & 200 & 0.027977 & 0.61830 & 0.147 \\
& FLT1  $|$ C13S523   & 200 & 0.066714 & 0.64997 & 0.147 \\
& FLT1  $|$ C13S524   & 164 & 0.004304 & 0.62223 & 0.147 \\
& KDR   $|$ C4S1877   & 145 & 0.000717 & 1.07706 & 0.111 \\
& KDR   $|$ C4S1878   & 101 & 0.164993 & 0.13573 & 0.111 \\
& KDR   $|$ C4S1884   &  95 & 0.020803 & 0.29558 & 0.111 \\
& VEGFA $|$ C6S2981   &  69 & 0.002152 & 1.20645 & \\
& VEGFC $|$ C4S4935   &  91 & 0.000717 & 1.35726 & \\
\midrule
Q2 &&&&& 0.008 \\
& BCHE  $|$ C3S4869  &  54 & 0.000717 & 1.01569 & 0.001 \\
& BCHE  $|$ C3S4875  &  59 & 0.000717 & 1.09484 & 0.001 \\
& LPL   $|$ C8S442   &  69 & 0.015782 & 0.49459 & \\
& SIRT1 $|$ C10S3048 &  54 & 0.002152 & 0.83224 & 0.330 \\
& SIRT1 $|$ C10S3050 &  72 & 0.002152 & 0.97060 & 0.330 \\
& VNN1  $|$ C6S5380  & 138 & 0.170732 & 0.24437 & \\
& VNN3  $|$ C6S5441  &  59 & 0.098278 & 0.27053 & 0.066 \\
& VNN3  $|$ C6S5449  &  57 & 0.010043 & 0.66909 & 0.066 \\
\bottomrule
\end{tabular}
\label{tab:snplist}\\

\end{table}

\begin{table}[h]
\caption{Proportion of common and rare variants of the true SNPs
found among the top 100 SNPs.}
\centering
\begin{tabular}{l rrrr rrr}
\toprule
Q1 \\
& Proportion (\%)  & CAR  & COR  & NEG  & MCP  & BOOST & LASSO \\
&    Common        & 0.56 & 0.71 & 0.63 & 0.74 & 0.71  & 0.73 \\
&    Rare          & 0.44 & 0.29 & 0.37 & 0.26 & 0.29  & 0.27 \\
\midrule
Q2 \\
& Proportion (\%)  & CAR  & COR  & NEG  & MCP  & BOOST & LASSO \\
&    Common        & 0.28 & 0.41 & 0.36 & 0.44 & 0.42  & 0.43  \\
&    Rare          & 0.72 & 0.59 & 0.64 & 0.56 & 0.58  & 0.57  \\
\bottomrule
\end{tabular}
\label{tab:props}\\

\end{table}

In \tabcite{tab:snplist} we list the SNPs identified by the CAR score among the top 100 SNPs
in at least 50 of 200 repetitions along with their 
minor allele frequency (MAF) and BETA values.
We consider SNPs with a MAF value smaller than 0.01 as rare and SNPs
with a larger MAF value as common variants.
The BETA value measures the effect size in the actual
simulation of the phenotype \citep{AD+2011}.
We find large differences between true positive SNPs of the two phenotypes. 
Whereas Q1 is characterized by SNPs with strong effects and moderate MAFs,
the true SNPs for Q2 have a very low MAF and are  much harder to detect.
Interestingly, most of the SNPs recovered by CAR scores are rare SNPs
with comparatively large BETA values. 
Common SNPs are found as well, then also with small effect values.
Thus,  CAR scores are successful in achieving a high
true positive rate because they not only allow to identify common SNPs 
 but also SNPs with small MAF if a strong signal is present (large BETA).

The last column in \tabcite{tab:snplist} provides information about the average absolute
correlation among all true SNPs for Q1 and Q2 as well as  among the identified SNPs
on the same gene.
We observe that the true positive SNPs in Q1 best identified 
by the CAR score are highly correlated within the same gene.
This demonstrates that the CAR score successfully utilizes the correlation 
structure among SNPs to optimize the ranking.
For phenotype Q2 the correlation among the true SNPs is generally lower
compared to Q1, still except for BCHE the correlation among  SNPs on 
the same gene is larger compared to the average correlation 
between a randomly chosen pair of true SNPs.

Finally, in \tabcite{tab:props} we provide the proportion of rare and common SNPs
found among the top ranking 100 SNPs for each methods.  This also shows that
the proposed approach based on CAR scores is effective in finding rare SNPs.

%% file: 4-discussion.tex
\section*{Conclusions}
 
Large scale simultaneous SNP selection is a statistically and computationally
very challenging task.  
To this end, we have introduced here a novel algorithm based on CAR score regression
that can be applied effectively in high dimensions.
Subsequently, in a comparison study we have  investigated five multivariate
regression-based SNP selection approaches with regard to their ability
to correctly recover causal SNPs and corresponding SNP rankings.

As overall best method we recommend  using CAR
scores since this method was the only approach not only consistently outperforming
the competing other multivariate SNP selection procedures
in terms of identified true positives but also the only approach uniformly improving
over simple univariate ranking by marginal correlation.
In addition we have shown that CAR scores also are successful in detecting rare variants
which recently have been recognize to be important indicators for human disease \citep{BB2008,MK2010}.

%% file: 5-acknowledgments.tex
\section*{Acknowledgements}

We thank Peter Ahnert, Arndt Gro\ss, Holger Kirsten, Abdul Nachtigaller, and Markus Scholz  for helpful discussion.  
Part of this research was supported by BMBF grant no. 0315452A (HaematoSys project).
The Genetic Analysis Workshop is supported by NIH R01 GM031575.
Preparation of the Genetic Analysis Workshop 17 simulated exome data set was supported in part by NIH R01 MH059490 and used sequencing data from the 1000 Genomes Project (\url{http://www.1000genomes.org}).